# Competition and extinction explain the evolution of diversity in American automobiles

**Erik Gjesfjeld*[1], Jonathan Chang[2] Daniele Silvestro[3], Christopher Kelty[4], Michael Alfaro*[2]**

[1] Institute for Society and Genetics, University of California, Los Angeles
[2] Departments of Ecology and Evolutionary Biology, University of California, Los Angeles
[3] Department of Ecology and Evolution, University of Lausanne, Lausanne, Switzerland
[4] Department of Anthropology, University of California, Los Angeles

*Corresponding authors

## Abstract

One of the most remarkable aspects of our species is that while we show surprisingly little genetic diversity, we demonstrate astonishing amounts of cultural diversity. Perhaps most impressive is the diversity of our technologies, broadly defined as all the physical objects we produce and the skills we use to produce them. Despite considerable focus on the evolution of technology by social scientists and philosophers, there have been few attempts to systematically quantify technological diversity and therefore the dynamics of technological change remain poorly understood. Here we show a novel Bayesian model for examining technological diversification adopted from paleontological analysis of occurrence data. We use this framework to estimate the tempo of diversification in American car and truck models produced between 1896 and 2014 and to test the relative importance of competition and extrinsic factors in shaping changes in macroevolutionary rates. Our results identify a four-fold decrease in the origination and extinction rates of car models and a negative net diversification rate over the last thirty years. We also demonstrate that competition played a more significant role in car model diversification than either changes in oil prices or gross domestic product. Together our analyses provide a set of tools that can enhance current research on technological and cultural evolution by providing a flexible and quantitative framework for exploring the dynamics of diversification.

**Introduction**
Despite the ubiquity of technology in our society, there remains no widely accepted view or quantitative framework for understanding technological change[1]. Rather, the subject is characterized by a series of generic frameworks[2,3] and unrelated case studies of successful or popular designs.[4-7] This myopic focus on "success stories" poorly characterizes the dynamic processes that shape cultural diversity through time and does not provide a framework for evaluating alternative hypotheses for successes or failures. We argue here that the field of macroevolution provides novel perspectives and tools for understanding the dynamics of technological diversification. Our research introduces a flexible statistical framework adopted from paleobiology that provides for the quantification of evolutionary rates of cultural change through time.

**Macroevolution, Technology and Diversity**
Evolution has been and continues to be a valuable source of methods and theories for the study of human culture. Previous research has demonstrated that human culture undeniably evolves, but to what degree cultural change mirrors biological change remains an unsettled question[8]. The evolution of technology is a topic in which the evolutionary analogy has been particularly contentious, with debate often centered on the unit of evolutionary analysis, the replication of technological designs and the applicability of branching models to understanding the evolution to intentionally designed objects. This paper presents an alternative perspective to the study of technological evolution that highlights the concept of diversity and a suite of macroevolutionary methods useful in quantifying the dynamics of technological diversification.

In this research, technological diversity is conceptualized as the number of different technological lineages represented in a system. This definition of diversity is different from disciplines that acknowledge diversity as having the additional dimensions of *balance* and *disparity*[9], but is analogous to the concept of species richness in biology, where a large number of methods are available for characterizing this component of diversity through time. The emphasis on the *richness* or *variety* of technological lineages is not to diminish the importance of *balance* and *disparity*, but rather to acknowledge that the information necessary to examine these components of diversity is not often available for historical or complex technological systems. For example, in order to evaluate *balance*, it is necessary to have an accurate estimate of the abundance of technological products over multiple time periods, which is often unavailable or unreliable when using historical or archaeological data. Alternatively, the calculation of *disparity* requires the detailed knowledge of product characteristics to distinguish between product classes, an arduous task for complex technologies.

To gain deeper insight into the diversification dynamics of technological systems, this paper presents a flexible set of macroevolutionary methods for examining the rate of diversification of technological linages through time. This methodology has several advantages to existing frameworks. First, diversity (variety) in our approach is determined by both the rate of origination and extinction of technological lineages. This is a more comprehensive representation of diversity than studies of technological evolution that analyze only the emergence of specific or successful designs. Second, our approach explores diversification dynamics in continuous time through the implementation of a birth-death model. One advantage of the birth-death framework for cultural and technological systems is that the process does not require the construction of evolutionary histories connecting material objects or the construction of detailed lists of product traits. The central assumption of this birth-death process is simply that novel variants emerge from the pool of existing variation in the previous time period. Third, our method can be applied to any dataset in which the dates of origination and extinction are known or only partially known. Our method is fully integrated into a Bayesian modeling framework allowing for the testing of competing hypotheses through the comparison of Bayes Factors. Through the implementation of our method we are able to 1) identify a continuous rate of technological origination and extinction through time, 2) compare the effects of intrinsic and extrinsic factors on rates of technological diversity and 3) explore the future rate of technological diversification.

**American Automobiles**
We investigated the diversification of American commercial automobiles following their origin in the 19$^{th}$ century to present. Automobiles are one of the most iconic examples of American mass technology in the 20$^{th}$ century and represent a successful, enduring and widespread human technology that is exceptionally diverse. Using the rich historical record of American automobiles, we focus our analysis at the level of the car model as models have a commercial and cultural reality that persists through time despite generational changes in features and physical appearances (Fig. 1). Because human technology, including cars, often includes an internal classification of diversity (we create both things and the categories for things), we rely on self-designated car model names and production years to distinguish product lineages. These "inherent" car model classifications is a different and more flexible approach than is commonly used in cultural evolutionary studies, where the expert attribution of traits, the categorization of objects or designation of age is something that a scientist does independently and often according to a set of assumptions about what traits are most important.

In addition to estimating the rate of car model diversification through time, we hypothesized that several continuous time factors might influence car model diversity. Cars are expensive consumer goods and we expected that a higher variety of car models would be manufactured during periods of high economic growth and a lower variety of models would be produced during lower economic growth. To create a proxy for economic growth we used the yearly change in overall gross domestic product from 1896 to 2014. Additionally, shifting oil prices have led to government-regulated changes in car manufacturing priorities, so we reason that changes in oil prices through time might also drive changes in car model diversity. Finally, we hypothesized that car model diversity may be influenced by more intrinsic factors such as the number of other car models that were also on the market. In this scenario, the number of other car models serves as a proxy for competition between car models and can act as a constraint on the overall variety of cars models present in a given year (diversity dependence). We further explore the influence of competition on existing car models by estimating the future diversity of gasoline powered and alternative fuel vehicles within a diversity-dependent market and an unbounded market. It is important to note that our combination of extrinsic (GDP and Oil Price) and intrinsic (competition) represents a small sample of continuous or discrete variables that could be evaluated in our Bayesian framework.

**Methods**
*Data*
The first production year of each car model is considered as an origination event with the last production year characterized as an extinction event. The production year, make and model name of automobiles used in this research were obtained from the parts and accessories finder associated with Ebay Motors.[10] Using this database, the make, model and production year were obtained for cars and trucks (excluding semi-trailer trucks) produced in the United States between 1896 and 2014. In total, 3,575 different car models were identified with 172 unique manufacturers.

*Bayesian Estimation of Diversification*
The diversification of car models through time is modeled here as the product of both origination and extinction. The estimation of origination and extinction rates utilizes an underlying birth-death model, which assumes these processes are random events occurring through continuous time. The probability of events occurring at any given time is determined by parameters of the birth-death process, which express the expected number of origination (or extinction) events per lineage per units of time. Rates can vary through time by introducing rate shifts. [11-13]

This approach uses a birth-death Markov chain Monte Carlo (BDMCMC) algorithm to jointly estimate the number of rate shifts, their temporal placement and the origination and extinction rates. In contrast to paleontological datasets, we assume that the origination and extinction times of car models are known and do not need to be estimated as part of a preservation process.[14] BDMCMC runs consisted of 10,000,000 generations with samples taken every 5,000 generations. Efficient chain mixing and effective sample sizes were assessed using Tracer.[15] Origination and extinction rates were summarized for one-year time bins with mean values and 95% highest posterior densities (hpd) calculated. Posterior estimates of origination and extinction rates are displayed as rate-through-time plots with the difference between origination and extinction rates forming the net diversification rate. The most probable numbers of shift points are inferred from the BDMCMC sampling frequencies and the best fitting rate model.[11]

*Correlated Diversification*
To test hypotheses of correlated diversification between car models and these intrinsic and extrinsic factors, we collected data on GDP [16,17] and oil price[18] from 1896 to 2014. We also extracted the number of car models available each year (diversity) from the Ebay database for our study period. We implemented linear and exponential models for these continuous variables using the following formulas:
$$\lambda(\tau) = \lambda_0 + \lambda_0(\gamma_\lambda * \tau)$$
$$\lambda(\tau) = \lambda_0 * \exp(\gamma_\lambda * \tau)$$

where $\lambda_0$ is the origination rate at present, $\gamma_\lambda$ is the correlation parameter for origination, and $\tau$ is the covariate value through time (*i.e.* the number of car model lineages, price of oil, gross domestic product). The same equations with an independent $\gamma_\mu$ parameter were also implemented for inferring correlation with extinction rates at the present ($\mu_0$). Under each of these models we ran a MCMC analysis with 1 million iterations. The posterior estimates of $\gamma$ parameters indicate positive correlation ($\gamma > 0$), negative correlation ($\gamma < 0$) or no correlation ($\gamma \approx 0$). We estimated the fit of different models by their marginal likelihoods, quantified through thermodynamic integration, and used them to rank the three correlates (diversity, oil price, GDP) by their ability to explain rate variation in the data.

*Forward Simulations*
We estimated the expected diversity of car models for the next 15 years, using forward simulations under a birth-death process. In the first set of simulations we fixed the origination and extinction rates to their posterior estimate at the present resulting from the BDMCMC analysis, that is $\lambda_0 = 0.103$, $\mu_0 = 0.118$ (thus implying negative net diversification). We started the simulations with an initial set of 135 car models (the number of extant models as of 2014) and maintained constant origination and extinction rates through time. After generating 1,000 birth-death simulations we summarized the resulting diversity through time as mean and 95% confidence interval. We followed a similar procedure on a data set of 47 hybrid and electric car models produced worldwide since 1996 (we excluded few older models which represented too few data points for the analysis). In a comparison between constant rate and diversity dependent birth-death models, the constant rate model was weakly supported by Bayes factors (logBF = 5.392). The estimated origination and extinction rates at the present indicated a strongly positive net diversification: $\lambda = 0.253$, $\mu = 0.123$. We ran a first set of forward simulations using constant birth-death rates and an initial diversity of 28 car models (the number of extant models as of 2014). We then hypothesized that the constant rate model was favored by Bayes factors because origination and extinction rates have not yet changed in response to the increasing diversity of electric car models and that hybrid and electric car might undergo a similar diversity dependent process as American cars. Thus, we repeated the simulations using the origination and extinction rates estimated from hybrid and electric cars as initial rates, but transforming them exponentially as a function of diversity based on the $\gamma_\lambda$ and $\gamma_\mu$ parameters estimated from American cars (Table 1). We summarized the diversity resulting from 1,000 birth-death simulations through time as mean and 95% confidence interval.

**Results**
Results from our Bayesian analysis indicate that origination and extinction rates of American car models follow closely matching trajectories, with high rates early in their history followed by a gradual decline leading to lower rates in more recent history (Fig. 2). In addition to time averaged rates, our framework also allows us to identify historical periods that are most likely to be associated with significant rate changes or shifts in diversification rates. Our analysis suggests two shift points for the origination rate: 1933 corresponding to the Great Depression and 1984 coincident with the reengineering of cars to meet fuel efficiency standards.[19] Similarly, we identify two shift points for the extinction rate: 1935, once again corresponding to the Great Depression and 1960 which marks the height of the ``Big Three'' dominance in the US automotive market (see shift points in Fig. 2).

Competition and historical factors have both been identified as potential components in shaping diversity. Here we model two historical factors, oil price and gross domestic product (GDP), as continuous variables and compare them to a competition model, which defines competition as the number of existing car models available in any year. Results from this hypothesis testing strongly support diversity-dependent (competition) models over all other competing continuous-time models (Table 1) with exponential and linear versions of oil price and GDP models fitting the data much more poorly than either exponential or linear diversity models.

**Discussion**

One surprising outcome of this research is that since the 1980's, the extinction rate of American car models has been higher than the origination rate, leading to a negative net diversification rate (Fig. 3A). This indicates that over the last thirty years more American car models have been lost per year than gained. Perhaps equally surprising is that the negative net diversification rate coincides with a decrease in origination and extinction rates. This means that although fewer new car models have been introduced towards the present, the lifespan of these models has tended to increase (Fig. 3B).

The linked dynamics of declining origination and extinction rates through time with an attendant increase in the average longevity of surviving models is consistent with theories of technological evolution that predict the emergence of dominant designs.[20,21] As Abernathy and Utterback predicted[22], when technologies become increasingly specialized over time the cost of implementing innovations becomes progressively higher, leading to fewer radical innovations and reduced diversity. Our research is consistent with the idea that once "dominant" car models (or manufacturers) are established after World War II, short-lived "experimental" models become less common due to increased costs of producing these models relative to established designs, leading to an overall drop in net diversification. At this point, the focus of competition in the auto industry shifts from competition between designs to competition between cost as car models become less-differentiated by functional design.[23] It can be argued that in this more mature economic environment increasingly model longevity is an effective management strategy as it reduces risky investments in new car models, increases brand loyalty[24] and helps to stabilize the costs of production[25]. Therefore, despite numerous incremental innovations and explosive growth in emerging markets, our results demonstrate that the rate of car model diversification has continued to decline over the last 30 years. The implication of this finding is that the American automotive industry may be locked into the "long-tail" of its life-cycle[23] with the historical trajectory of the industry strongly constraining present car model diversity and the emergence of radical new car models and designs.

The overall pattern of diversity dependence is strong evidence that the diversification of car models has been shaped by competition.[26] Although automobile manufacturers are known to engage with one another in economic competition, this study quantitatively links diversity-dependence as the dominant control on the richness of objects in a technological system. This finding provides an important first step for future studies of technological evolution by providing an initial set of empirical and quantitative expectations for the mode and tempo of technological evolution. For example, by assuming diversity-dependence of the same intensity as gas-powered American car models, as provided by our Bayesian model, we predict that electric and hybrid cars may be experiencing the early stages of a radiation event, with dramatic diversification expected in the next three to four decades (Fig. 4). However, without accounting for the role of competition between car models, the diversity trajectory of electric and hybrid models may initially appear unbounded and produce inaccurate estimates for how car models and technologies diversify through time.

**Conclusions**

Just as the fossil record provides evidence for biological change through time, the archaeological and historical record has an important role to play in our understanding of technological and cultural evolution by providing empirical evidence for change through time. Our analyses of American car models reveals the shifting roles that origination and extinction have played in shaping diversity in one of the most important and ubiquitous technologies of the 20[th] century. Furthermore, we demonstrate how the analysis of cultural change in a birth-death framework provides a means for testing alternative hypotheses about the extrinsic and intrinsic controls on technological diversification. Our approach is flexible and easily adapted to other cultural systems where a record of first and last appearances of artifacts is available. Overall, the quantitative study of diversification within a macroevolutionary framework offers enormous potential to enrich our understanding of cultural and technological change.

**Data Availability**

Data on the make, model and production year of cars is accessible at the Ebay Car and Truck Parts and Accessories Finder (http://www.ebay.com/motors/Parts-Accessories/). Automobile makes, model names and production years for American, global plug-in hybrid electric vehicles (PHEV) and battery electric vehicles (BEV) are available at https://github.com/erikgjes/cars and in the Dataverse repository (http://dx.doi.org/10.7910/DVN/GPLE2D). Data on Gross Domestic Product and oil prices were derived from the following resources:

www.bea.gov/national/index.htm#gdp (Gross Domestic Product)
http://measuringworth.org/usgdp (Gross Domestic Product)
www.eia.gov (Oil Prices)

The PyRate package (including PyRate Continuous) used to implement the Bayesian analyses with detailed manual and template input files is accessible at http://sourceforge.net/project/pyrate/.

**Tables**

| Covariate | Model | $\lambda_0$ | $\mu_0$ | $\gamma_\lambda$ | $\gamma_\mu$ | Marg. Likelihood | BF |
|---|---|---|---|---|---|---|---|
| Diversity | Exponential | 0.260 | 0.247 | -2.307 | -2.147 | -17582 | - |
| Diversity | Linear | 0.290 | 0.274 | -1.621 | -1.572 | -17626 | 43 |
| Oil Price | Exponential | 0.024 | 0.053 | -2.815 | -1.719 | -18386 | 803 |
| Oil Price | Linear | 0.040 | 0.076 | -5.432 | -2.038 | -18540 | 958 |
| GDP Change | Exponential | 0.206 | 0.199 | -1.779 | -1.831 | -18704 | 1121 |
| GDP Change | Linear | 0.200 | 0.191 | -0.896 | -0.814 | -18800 | 1217 |

**Table 1.** Model fit quantified as marginal likelihoods for three covariates under analysis with Bayes factors providing support for the best model (exponential density dependence) against all alternative models. Model parameters include the baseline origination and extinction rates ($\lambda_0, \mu_0$) and the correlation parameters for origination and extinction

**Figures**

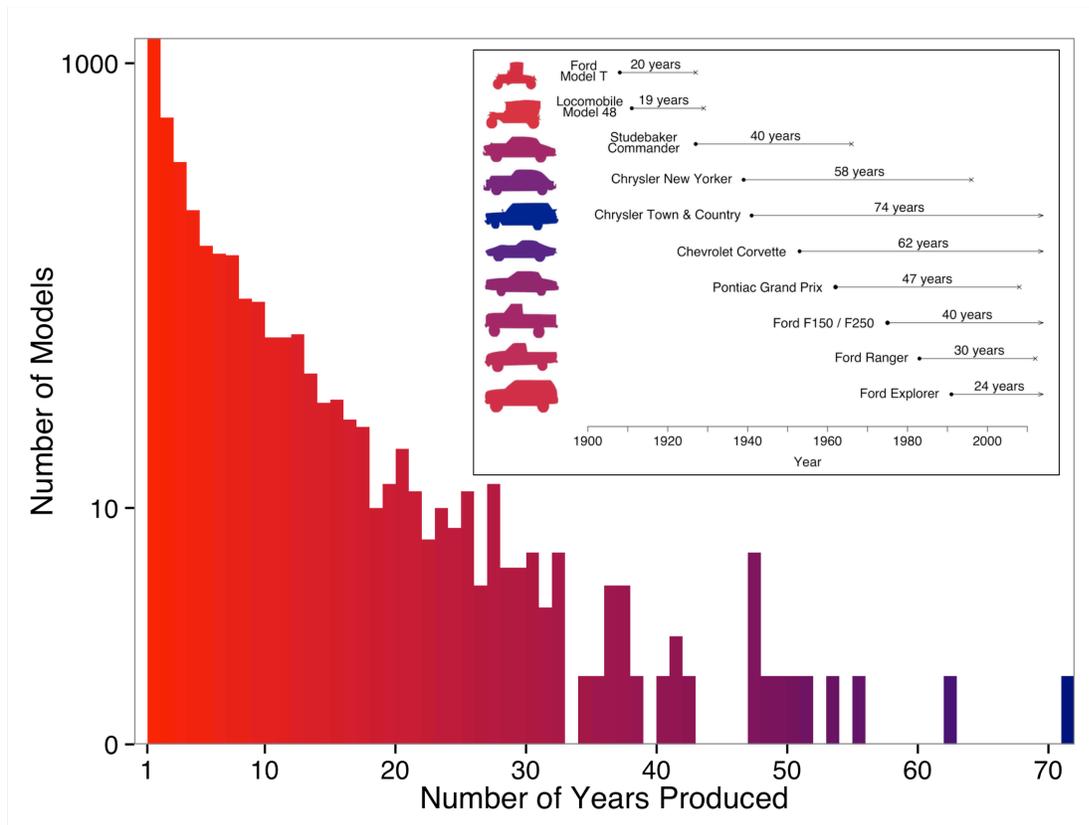

**Figure 1.** Histogram showing the distribution (one-year bins) of the number of years each car model in our data set was sold for (model lifespan). Inset shows the longest produced car model that originates in each decade of the 20th century with first production year indicated by a filled circle, last production year indicated by an x and models still currently in production indicated by an arrow. Colors of cars in the inset is representative of their bin in the histogram.

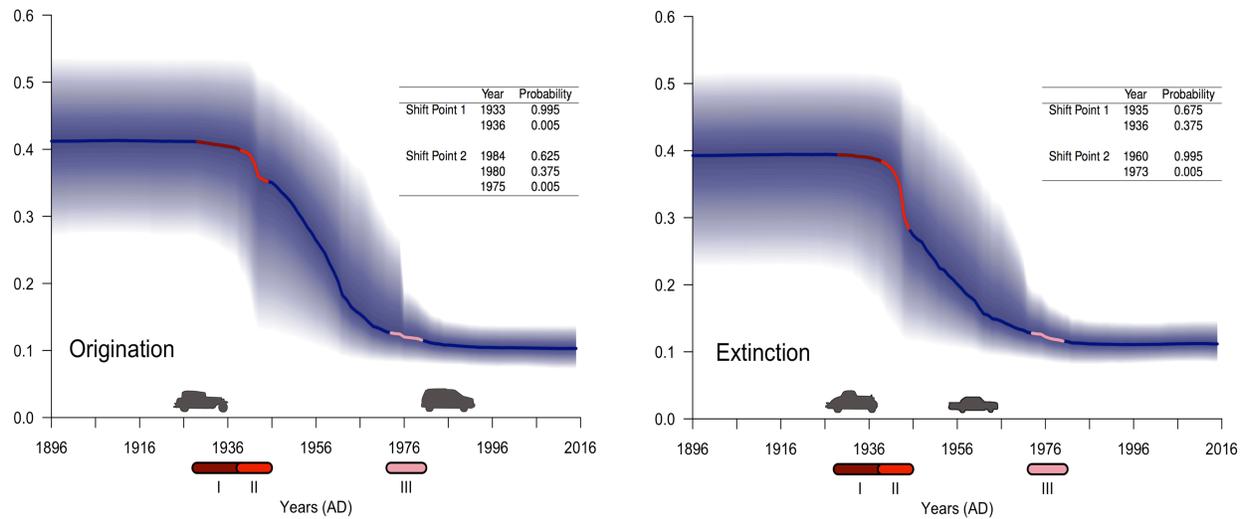

**Figure 2**. Rates of origination and extinction through time for car models. Solid lines indicate the mean origination and extinction rate for each year with shaded areas representing the 95% highest probability densities. Tables provide sampling frequencies for shift point probabilities. Car silhouettes represent models that either originated (1933 Nash Ambassador and 1984 Dodge Caravan) or went extinct (1935 LaSalle Series 50 and 1960 Edsel Ranger) at the estimated shift points. Red shaded areas represent significant global economic events that are commonly thought to influence car diversity including the Great Depression (I), World War II (II) and the Arab Oil Crisis (III).

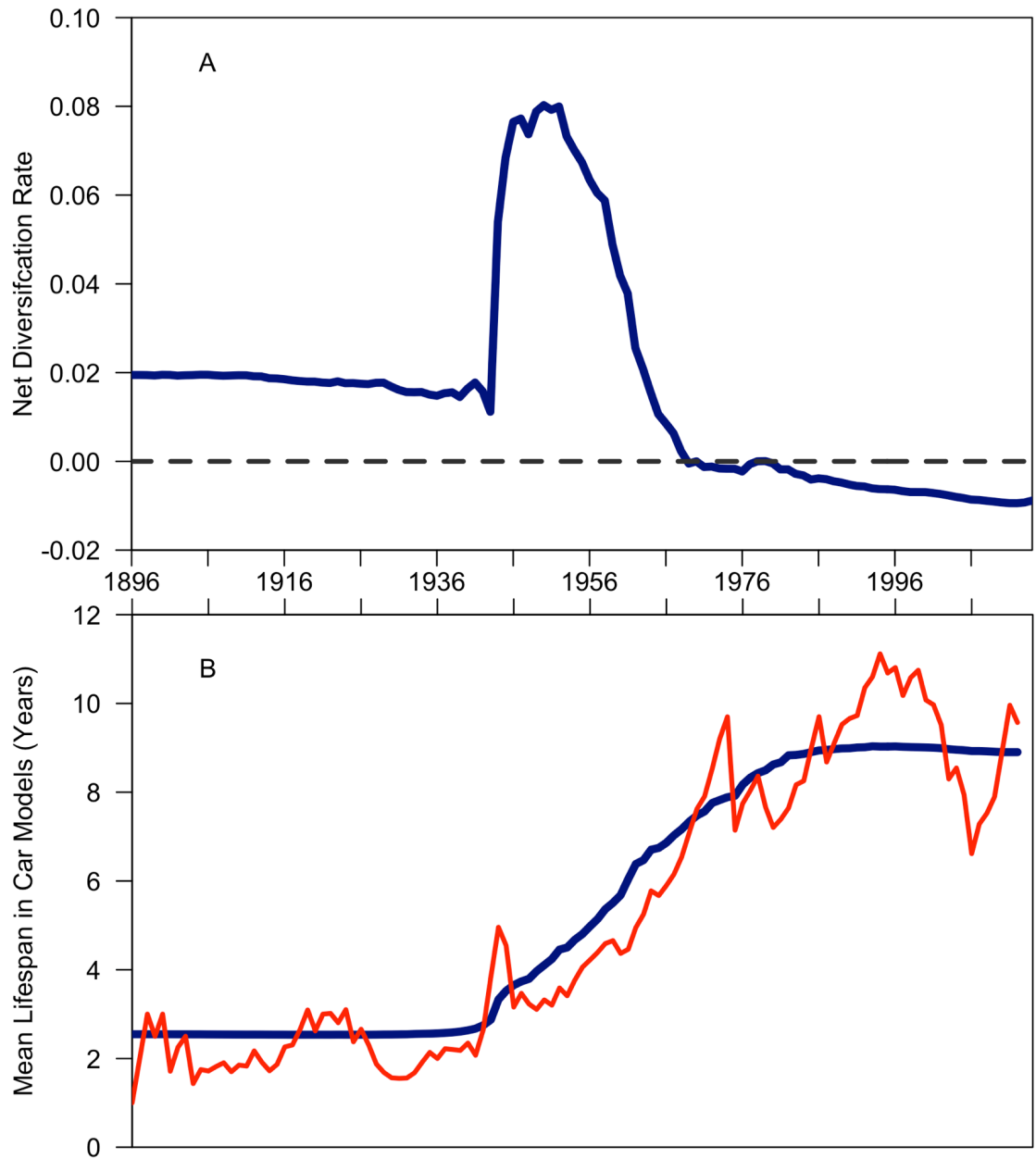

**Figure 3.** Plots showing changes over time in net diversification (A) and mean lifespan of car models (B). Net diversification rate of car models is equal to the origination rate minus the extinction rate with a dotted line indicating a net diversification of zero. The empirical mean lifespan of car models is shown in red while the expected mean lifespan ($\mu^{-1}$) based on our data is provided in navy blue.

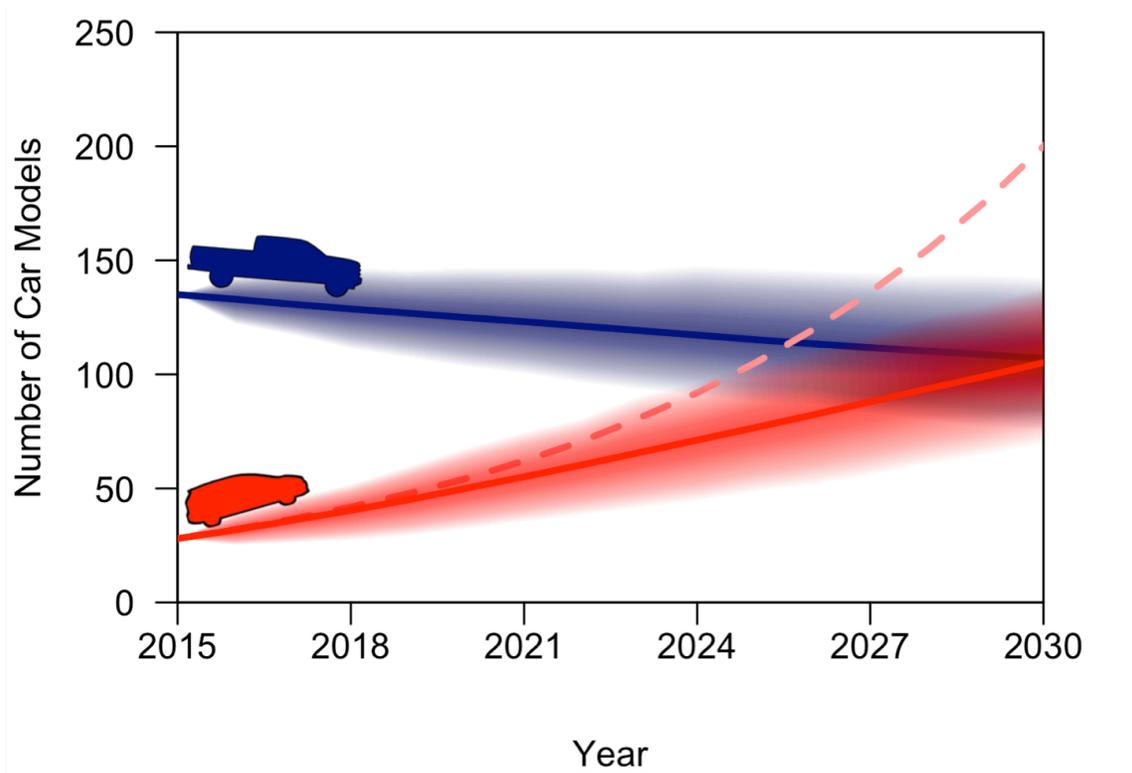

**Figure 4**. Projected diversity of American gas powered automobiles (navy blue) as well as battery electric and plug-in hybrid vehicles at constant growth (dotted pink) and diversity-dependent growth (solid red) with 95\% confidence intervals (shaded areas)